\documentclass[epj]{svjour}
\usepackage{graphics}
\usepackage{graphicx}
\usepackage{epsf}
\usepackage{bm}
\usepackage{amssymb}
\usepackage{amsmath}
\usepackage{subfigure}
\begin{document}
\title{Density, phase and coherence properties of a low dimensional Bose-Einstein systems moving in a disordered potential}
\author{Chaitanya Joshi  \inst{}, Sankalpa Ghosh \inst{}
}
\institute{Department of Physics,
Indian Institute of Technology, Delhi,
Hauz Khas,
New Delhi 110016,
India}
\date{Received: date / Revised version: date}
\def \beq{\begin{equation}}
\def \eeq{\end{equation}}
\def \bea{\begin{eqnarray}}
\def \eea{\end{eqnarray}}
\def \bem{\begin{displaymath}}
\def \eem{\end{displaymath}}
\def \P{\Psi}
\def \Pd{|\Psi(\boldsymbol{r})|}
\def \Pds{|\Psi^{\ast}(\boldsymbol{r})|}
\def \Po{\overline{\Psi}}
\def \bs{\boldsymbol}
\def \bl{\bar{\boldsymbol{l}}}

\abstract
{We present a detailed numerical study of the dynamics of a disordered
one-dimensional Bose-Einstein condensates in position and momentum space. We particularly focus on the region where non-linearity and disorder simultaneously effect the  time propagation of the condensate as well
as the possible interference between various parts of the matter wave. We report oscillation between spatially extended and localized behavior for the propagating condensate which dies down with increasing non-linearity. We also report intriguing behavior of the phase fluctuation and the coherence properties of the matter wave. We also briefly compare these behavior with that of a two-dimensional condensate. We mention the relevance
of our results to the related experiments on Anderson localization and indicate the possibility of future
experiments}


\maketitle

\section{Introduction}
The spatial behavior of the time dependent density profile of a cold atomic Bose-Einstein condensate (BEC)
in presence of a optical disorder potential has been  under intense experimental \cite{disorder1},
\cite{disorder1a}, \cite{disorder3}, \cite{sch05} and theoretical investigation \cite{GC04}, \cite{modugno}, \cite{Palencia1}, \cite{Palencia2}, \cite{shapiro1}, \cite{AGM}in recent years. Such interest in this issue is attributed to the possibility of observing the phenomenon called Anderson localization (AL) \cite{Ander58} of matter waves. AL corresponds to the spatial localization of  a wave due to destructive interference between multiply scattered waves from the modulations of a disorder potential. The  BEC of ultra cold atoms is considered as an  ideal system to observe such localization. This is  because of the system forms a giant matter wave of ultra cold bosonic atoms with very  weak interaction among them. Additionally the system can be prepared in reduced dimension  and according to the scaling theory of localization \cite{Gang479} the effect of disorder gets enhanced in that case. In two recent experiments carried out
in Orsay \cite{ALobservation1} and in LENS \cite{ALobservation2} Anderson localizations of matter wave were observed
respectively in a weakly interacting quasi one dimensional {\bf repulsive} condensate and non-interacting one dimensional condensate. The rapid progress in this field also has been recently summarized in two recent reviews by Orsay and LENS group \cite{BECdisorderreview1}, \cite{BECdisorderreview2}.

This paper investigates the the joint effect of non-linearity and disorder on the time evolution of the co-ordinate and momentum space density, phase fluctuation and coherence properties of such condensate
We begin with the study of time dependent density distribution in co-ordinate as well as in momentum space. According to the scaling theory of localization \cite{Gang479} in one spatial dimension the motion of a quantum mechanical particle is either ballistic, or localized in presence of disorder with infinitesimal strength, and the diffusion is absent. However recent analysis \cite{Palencia2},\cite{shapiro1},\cite{BECdisorderreview1} showed that the presence of interaction changes
this situation and localization properties become momentum dependent.
Different wave-vector components
of the propagating Bose-Einstein condensates will behave differently as they get scattered from the random modulations of the disordered potentials. And
this motivates us to numerically investigate the localization properties of the fast as well as slow components of the static condensate.
Additionally we study the phase fluctuation and the associated coherence of the condensate as a function of time which provides new and interesting informations about this system. In lower dimensional systems  strong phase fluctuations quite often destroys the long range phase coherence in Bose-Einstein systems. Though it was  shown \cite{PSW00} that for a low dimensional condensate the phase coherence exists over  a length scale comparable with the typical system size, for the time evolved condensate in presence of a disordered potential this argument cannot be automatically extended. This motivates us to investigate the coherence properties of such time evolved condensate at least numerically.

To place our analysis in the experimental context let us mention that the issue addressed in most of the current experiments and the associated theoretical studies is to ascertain under what condition the system will manifest AL and what will be the impact of non-linearity on this localization process. The AL has been confirmed by the presence of an exponentially decaying envelope wave-function of the relevant system. In a recent experiment
\cite{ALobservation1}
the  time of flight image of the edge  of a quasi-one dimensional Rubidium  BEC which has low density and almost negligible interaction energy, showed this exponential decay.  Thus it has been concluded as
Anderson localized. In this experiment both the strength of the disorder as well as the density of the condensate are kept at a relatively lower value to achieve the regime of AL.
These experimental conditions are borne out of the fact that a very strong disorder  classically localizes a wave purely on energetic ground whereas for a high density condensate cloud the interaction effect is too strong to allow the Anderson localization to happen.
The preceding theoretical analysis \cite{Palencia2}, \cite{shapiro1} which has identified these experimental conditions suitable for the observation of Anderson localization can be summarized as follows. A typical trapped condensate
contains a range of momentum components. The higher momentum components that mostly reside at the low density edge of the condensate moves much faster as the condensate propagates through the disorder potential. After an initial stage of expansion in which most of the interaction energy of the condensate
gets converted into kinetic energy, these higher momentum components of the condensate are mostly present at the edge of the expanded condensate and behave like plane wave. These are almost free from the non-linear effect and thus
are appropriate for observing AL.

This should be contrasted to the behavior of the central part of the time evolved condensate which is mostly populated by the low momentum components
and contains non-linear effect.  The time evolved density profile of this central region nevertheless gets strongly modified by the disordered potential \cite{BECdisorderreview1}. However, why such modifications
do not lead to AL is still not completely understood. On the other hand it is the behavior of this section of the cloud which manifests the interplay between the non-linearity and the disorder and sheds light on the impact of the non-linearity on the AL. Our work address this unresolved issue by 
trying to find out the relative impact of non-linearity and disorder by studying numerically the time evolution of 
density modulation, phase modulation and coherence properties 
which combines the effect of density and phase. Thus this analysis provides us important information about the fate of interference induced 
quantum localization in presence of non-linearity and augment a number of theoretical studies that studied these issues previously \cite{modugno},\cite{Palencia2},\cite{shapiro1},\cite{AGM}.

\section{The Theoretical framework}
In this section we discuss briefly  various details of the theoretical framework of the problem. First we talk about the model disorder potential. Then we shall discuss the disordered non-linear Schr$\ddot{o}$dinger equation which will be solved to study the behavior of the disordered condensate

\subsection{Model of the disorder potential}\label{sec:disorder}
The results that we shall show in the subsequent sections have been generated with the disorder potential as described in the work of Huntley \cite{huntley} and has been used in a number of experimental
work to model the disorder potential. Here, the real and imaginary part of complex amplitude of light are constructed by Gaussian random numbers. Such a disorder potential is known as the speckle potential and the height of the speckle which characterizes the strength of the disorder potential is proportional to the standard deviation of the disorder potential around an almost zero mean value.
We have compared the results obtained in presence of such disordered potential with ones where we have created the disordered potentials simply from a uniformly distributed random numbers \cite{AGM} or real random numbers with Gaussian distribution. The later type of  random potentials have non-zero mean and its strength is  characterized by the first as well as the second moment of the distribution of the random variables. The details of the time dependent as well as stationary density profiles of condensates indeed differ from one realization of the disorder potential to another, but the basic features related to the localizations remained unaltered. All the results presented here are for single realization of the disorder potential. But repeating the calculation for various realizations of the disorder potential we found that the basic conclusions, these results deliver, do not depend on the details of a realization.

Also, the mean field description of the disordered condensate through the disordered Gross-Pitaevskii equation is valid ( for a more detailed explanation, see \cite{BECdisorderreview1})
as long as the
disorder potential fluctuates over a length scale that is larger than the so called healing length of the condensate. To ensure that this condition is obeyed
we discard from the Fourier spectrum of the disorder potential all wavenumbers that are above a given cutoff $k_c = 2 \pi / \lambda_c$ \cite{AGM}.The inverse Fourier transform provides a random potential that varies on length scales larger or equal to $\lambda_c$ and which can be formally written as
\beq V_d (x) =  \int dk e^{ikx}\left[
e^{-\left({\frac{k}{k_c}}\right)^M} \int d\zeta \, \ V_{s} (\zeta) e^{-ik\zeta}\right]\; ,
\label{disorder} \eeq
where $V_{s}$ is the original speckle potential.
In the final calculation we have used this smoothened disorder potential.
\subsection{Mean field description of the disordered condensate}\label{framework}
We have described the time evolution of the quasi-one dimensional {\bf repulsive} condensate through the mean field Gross-Pitaevskii (GP) equation. Most of the  recent experiments \cite{disorder1,disorder3,sch05,ALobservation1}
can be well explained by such mean field theory. From the pure theoretical perspective this also helps to address the issue
if one dimensional Nonlinear Schr$\ddot{o}$dinger equation (NLSE) shows Anderson localization
in presence of a disorder potential \cite{nlseal}, that its linear counterpart does.

Mean field GP equation is obtained from the microscopic description of the ultracold bosonic systems 
by replacing the bosonic field operator by its vacuum expectation values which is taken as condensate order 
parameter.  The details are given in a large number of text books ( for example, see ref. \cite{pita})
Typically for these
ultra cold atoms the interaction is extremely short ranged and characterized by a single parameter which
is the $s$-wave scattering length. In the GP equation the interaction is characterized by a  nonlinear term that is cubic in the order parameter. 


Our parameters are chosen such that
we are close to the typical experimental regime \cite{ALobservation1}. A quasi one dimensional condensate is
achieved by tightly trapping the condensate in transverse direction such that the only quantum mechanical
state occupied in the transverse direction is the ground state of a two-dimensional harmonic oscillator.
The mean field Gross Pitaevskii equation that describes such one dimensional condensate in presence of a speckle potential $V_{s}$ is given by \cite{PSW00}
\beq i\hbar \frac{\partial \Psi}{\partial t}=
-\frac{\hbar^2}{2m}\frac{\partial^2 \Psi}{\partial z^2}+
\frac{1}{2}m\omega_z^2 z^2\Psi + V_{s} \Psi + g_{1d} |\Psi|^2
\Psi \label{nlse1}\eeq
where $g_{1d}=2a \hbar \omega_{\perp}$ is the one dimensional effective coupling constant \cite{Olshanii98}
that depends on the $s$-wave scattering length $a$ and the radial harmonic trapping frequency $\omega_{\perp}$ of the condensate.  Here $\omega_z$ is the trapping frequency in the longitudinal direction. 
We first rewrite the Gross-Pitaevskii equation in a dimensionless form in the same way as in \cite{AGM}.

It becomes
\beq i\frac{\partial \psi_{w}}{\partial \tau} = -\frac{1}{2}\frac{\partial^2 \psi_{w}}{\partial x^2} +
\frac{1}{2}x^2 \psi_{w} + V_{d} \psi_{w} + gN|\psi_{w}|^2 \psi_{w}
\label{NLSE} \eeq
by choosing $\hbar \omega_z$ as the dimension of energy, $\sqrt{\frac{\hbar}{m \omega_z}}$ as the unit of length, $\omega_z^{-1}$
as the unit of time. $V_d$ is defined in Eq. \ref{disorder}. Since we shall be considering the time evolution of the condensate density after its release from the trap we set the trapping potential  $\frac{1}{2}x^2$ to zero.
We also denote the dimensionless co-ordinate $\frac{z}{\sqrt{\frac{\hbar}{m \omega_z}}}$ as  $x$ from now on where as for the time we set $\omega_z t = \tau$. $\psi_{w}$ that gives
the condensate wave function in presence of disorder is normalized to unity in the above equation and
the interaction coupling is thus proportional to the number of particles $N$.
As discussed in the section \ref{sec:disorder}, the
validity of this mean field equation in the presence of the disorder potential requires that the $\lambda_c > \xi$, where $\xi$ is the healing length of the condensate ( $\sqrt{\frac{1}{2gN}}$ from Eq. \ref{NLSE}).
In the following numerical calculations $\xi_{highest}=0.250$  for dimensionless chemical potential $\mu=6$, where as the $\lambda_c =1.25$, and,  satisfies the above criterion.

Using the parameters  for $^{87}Rb$  we found that under typical experimental condition
$g \approx .03$ \cite{ALobservation1}. The values chemical potentials $\mu$ (in the unit of $\hbar \omega_z$) that have been used  in the numerical analysis pertains to $\approx 10^3 - 10^4$ number of particles for a condensate. In this regime the Thomas-Fermi (TF) approximation works well
\cite{PSW00,AGM}
and this is also close to the typical experimental conditions.

We assume that at $t=0$ the condensate has a TF profile
for a given chemical potential $\mu$ \cite{AGM}. We have defined $L_{TF}=\sqrt{2\mu}$ as half of the size of such TF condensate. Thus the total number of particles $N$, that can be obtained by integrating TF density over this TF length, increases with increasing $\mu$. Since $N$ appears as
the coefficient of the non-linear term in eq. \ref{NLSE}, by increasing $\mu$ one actually increases the
strength of the non-linearity and make the condensate more interacting.

We shall obtain the condensate wave function and the corresponding density, $\rho_{w}(x,t) = |\psi_{w}(x,t)|^2$, at a later time $t$ by numerically integrating the above time dependent equation
(\ref{NLSE}) in the method outlined in \cite{BJM03} and has been used in \cite{AGM} over a discrete grid whose linear size is much larger than the $TF$ length of the condensate. We have also tried our numerical code for several grid sizes to make sure that the discretization process does not add 
any spurious feature
to our result. We will discuss more on the effect of numerical discretization in the next section with the help of results.
Additionally we have also used
an absorbing boundary condition so that any possibility of reflection from the boundary is eliminated. This describes the theoretical framework and the values of parameters that are used in our numerical analysis. In the following section we shall present our results and will analyze them.

\section{Results and discussions for one dimensional condensate}
\subsection{Time dependent density}
\begin{figure}[ht]
\centerline{ \epsfxsize 10cm \epsfysize 10cm
\epsffile{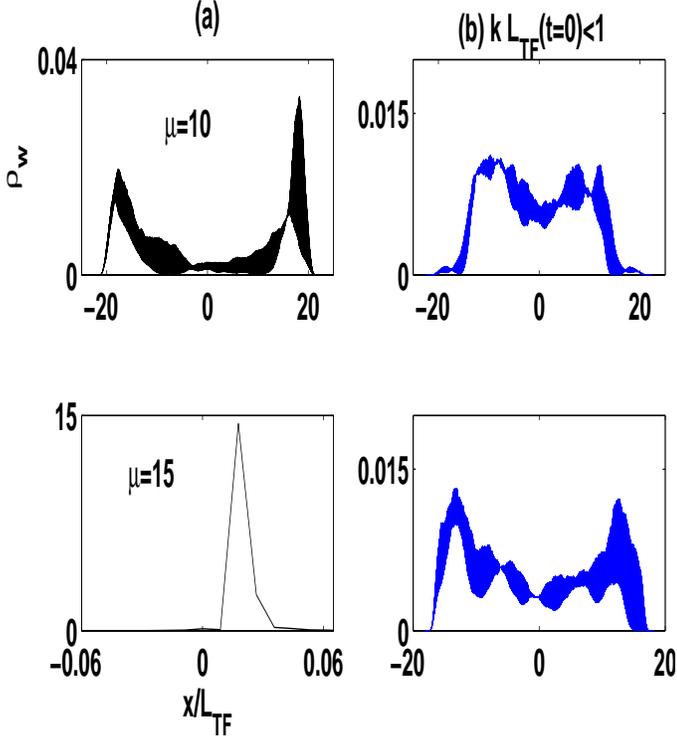}}
\caption{{\it color online}:{\em one dimensional condensate in a disorder potential. The disorder potential
has mean $0$ and standard deviation $0.59$. $\lambda_{c}$ for the disorder potential is $1.25$. All quantities are in the dimensionless unit.
(a)The left
column gives co-ordinate space density profile density profile
of an one dimensional condensate after time of flight of $\omega_{z} t=10$ for  $\mu=10$ and $\mu=15$. $x$ axis is in unit of TF length $L_{TF}$ of the static condensate in each case. The number of particles varies from $10^3$ - $10^4$ as the chemical potential changes (b) Corresponding time evolved density profile of the "slow condensate such that at $t=0$, $k L_{TF} <1$ are given in the right column for the same $\mu$}}
\label{CJSGfig1}
\end{figure}
\begin{figure}[ht]
\centerline{ \epsfxsize 10cm \epsfysize 10cm
\epsffile{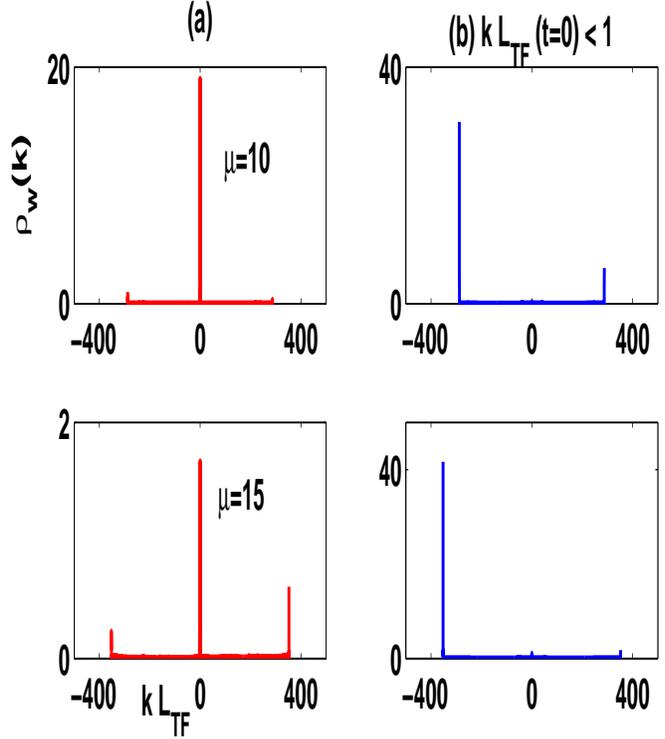}}
\caption{{\it color online}:{\em Momentum space profile of one dimensional condensate in a disorder potential. The disorder potential as well as other parameters are same as in the case of Fig. \ref{CJSGfig1}(a) The left column provides the momentum ($k$) space profile of the actual condensate.
(b) The right column provide the $k$ space density profile for the "slow condensate" }}
\label{CJSGfig1a}
\end{figure}

It has been shown  \cite{castingora} that for a propagating condensate, most of the interaction energy gets converted into kinetic energy over the time $ t \sim \frac{1}{\omega_z} \Rightarrow \omega_z t = \tau \sim 1$. Here we present the result at $\tau = 10$ in the Fig. \ref{CJSGfig1} and \ref{CJSGfig1a}. Each calculation have been performed for a number of realization of the disorder potential and features common to all such realizations are demonstarted 
graphically using one such representative realization. From time to
time, we shall also point out those features which may vary from one realization of the disorder potential to another.

All the subsequent figures broadly answer the following three questions :
\begin{enumerate}
\item How the density, phase and the coherence properties of the condensate evolves when the disorder potential is kept same but the strength of non-linearity ($\mu$)is varied ?
\item How the above properties evolve with time when the non-linearity is kept
same, but the strength of the disorder potential is varied ?
\item Finally, how the time evolution of all these properties changes when the higher Fourier components are removed from the initial profile of the condensate?This removal is done by only allowing the components satisfying $k L_{TF} <1$ at $t=0$ in the profile of the condensate where $L_{TF}$ is the TF length of the static condensate. Henceforth, to avoid repetition, in the rest of the discussion we shall refer this wave-number filtered condensate profile as {\bf slow condensate}.
\end{enumerate}
Whereas the importance of first two questions can be easily appreciated the relevance
of the last question demands some discussion. In reference \cite{shapiro1} for the case of a two-dimensional condensate with radially symmetric density profile it has been argued that that the components which satisfies the criterion $k L_{TF} < 1$ will show a diffusive a behavior after the initial
stage of ballistic expansion. In that case  $k$ is the absolute magnitude of the two dimensional wave vector and $L_{TF}$ corresponds to TF radius. In one dimension,
in the non-interacting limit
no diffusive motion can take place. However, it can be generally expected that the faster ($kL_{TF} >1$) and slower
component will response to the disorder differently.  Reference \cite{BECdisorderreview1} also describes
in detail how this leads to the existence of a mobility edge in one dimension in presence of non-linearity.  It is in this context the above mentioned numerical investigation has been carried out. We find that 
slowing down of the condensate in this way indeed effects its transport through the disorder.

The column $(a)$ of Fig. (\ref{CJSGfig1}) shows that density profile for the full condensate as the chemical potential is being increased for the same realization of the disorder potential. Unless otherwise stated in all the figures chemical potential $\mu$ is in the unit of
$\hbar \omega_z$. As pointed out, higher $\mu$
implies higher density of particles and thus higher strength of non-linear term.
The left column shows a strong localization as the non-linearity is increased by changing  $\mu=10$ to $\mu=15$.
The lower plot in column $(b)$ shows for the corresponding slow condensate shows that 
the removal of the higher $k$ density components from the initial profile resists this
strong localization at $\mu=15$. The patches that appear in the plots of the condensate density in  Fig. \ref{CJSGfig1} 
are actually very fine density modulations which can be seen by enlarging these portions of these figures.
The edge profile of all these disordered condensate shows a decaying envelope, where as bulk density modulation  does not have such decaying envelope for the bulk portion.

Given the fact that  we are considering a repulsive condensate, this strong localization
for higher non-linearity is a surprising result. However it may be noted that in the context of discrete non-linear Schr$\ddot{o}$dinger equation which is used to model a dilute BEC in presence of an additional  periodic potential
under the tight binding approximation, localized excitations due to both discreteness as well as  repulsive non-linearity
\cite{Smerzi01} was also reported. Thus one may argue whether this localization is an
artifact of the numerical discretization of the otherwise continuous functions in the GP equation. We will further discuss the possible 
effect of numerical discretization with the help of $k$-space density distribution in the next paragraph. 

To complement the above finding, in Fig. \ref{CJSGfig1a} we plot the $k$-space density
profile of the condensate for all the situations depicted in Fig. \ref{CJSGfig1}. For $\mu=10$, the momentum space density distribution of the time evolved condensate remains strongly  localized around  $k=0$ with two very small peaks at the corner of the first Brillouin zone. As it is increased to $\mu=15$, we see that density peaks occur at $k=0$ as well at  the two corners of the first  Brillouin zone. We note that we have considered in the numerical method
the $k$-space distribution within the first Brillouin zone only.
In the right side of the same figure, namely column (b), we see  that the removal of the faster components from the initial condensate density leads to a localization of the time evolved condensate density at different points along the $k$-axis which are far removed from  $k=0$ and 
are located near the first Brillouin zone corners. The scattering at the Brliiouin zone corners are a signature of Bragg scattering since 
$\vec{k}_{scatter} - \vec{k}_{initial} = \vec{G}$ where $G$ is the reciprocal  lattice vector of the lattice where the scattering takes place.
Here the only source of the periodic lattice is the numerical discretization. However numerical grid remains same throughout our calcualtion
whereas the $k$-space density $\rho_{w}(k)$ is not the same at the Brillouin  
zone corner for various values of $\mu$ as well as for the various strength  of the disorder potential. It also changes substantially for the slow condensate (
Fig. \ref{CJSGfig1a}, column (b) )
Moreover we checked that for other size of the numerical grid 
various features of the result remains almost same. Thus we may conclude that the density peaks at the corner of Brillouin zone  is 
partly due to the the numerical discreteness and the artificial lattice which leads to a strong Bragg scattering. Additionally, the peak also 
depends on the strength of non-linearity and disorder and also on the initial momentum distribution. However within the framework of this numerical 
investigation we are unable to find out to what extent the Bragg scattering is influenced by the later set of effects.

\begin{figure}[ht]
\centering
\subfigure[ {\em For $\mu=15$ the left
figure (a) gives co-ordinate space density in the absence of disorder for the condensate, obtained after 
the time of flight $\omega_z t = \tau =10$.
The right figure (b) gives the corresponding density distribution in the momentum space.}] 
{
    \label{CJSG2suba}
    \centerline{{ \epsfxsize 8cm \epsfysize 7cm \epsffile{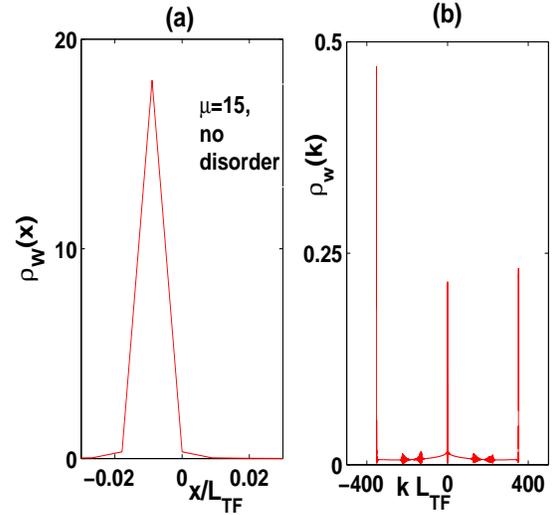 }}}
}
\subfigure[{\em Similar plots as in Fig. \ref{CJSG2suba} but for a stronger disorder  than one used in Fig. \ref{CJSGfig1}.The left
 figure (a)  gives the spatial density profile of the condensate. Plot (b) again gives the corresponding momentum space density.
The stronger disorder potential has  mean $0$, standard deviation is $1.15$ and 
the cutoff wavelength is $\lambda_c=0.31$. All other parameters are same as Fig. \ref{CJSG2suba}} ] 
{
    \label{CJSG2subb}
    \centerline{{\epsfxsize 8cm \epsfysize 7cm \epsffile{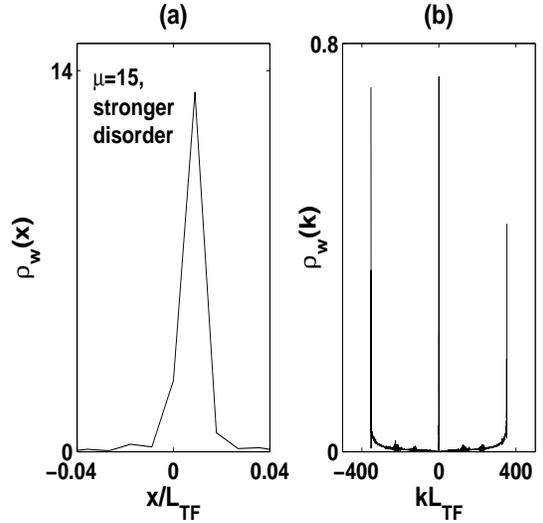 }}}
}
\caption{Time evolved density of the propagating condensate with and without disorder}
\label{CJSGfig2} 
\end{figure}
In Fig. \ref{CJSGfig2} we have compared the above mentioned density profiles with the case when
there is no disorder (Fig. \ref{CJSG2suba}) and when the  disorder potential is stronger (Fig. \ref{CJSG2subb}) for $\mu=15$.  
Comparison between left plot of Fig. \ref{CJSGfig1} (lower row) and  Fig. \ref{CJSG2suba} and Fig. \ref{CJSG2subb} 
shows that the strong localization of condensate density in co-ordinate space at $\mu=15$ is more 
due to the non-linearity (interaction) and not due to the disorder. 

The right plot of Fig. \ref{CJSG2suba}and Fig. \ref{CJSG2subb} again plots the corresponding 
$k$-space density distribution. The features agree with the Fig. \ref{CJSGfig1a} where a intermediate disorder strength is used with the showing of 
of density peaks at the corner of the first Brillouin zone apart of those in the center. As we have discussed the Brillouin zone density peaks 
is partly due to the Bragg refelction caused by numerical discretization, but its height also depends on strength of non-linearity, disorder as well
and changes for the slow condensate.

\begin{figure}[ht]
\centerline{ \epsfxsize 9.5cm \epsfysize 10cm
\epsffile{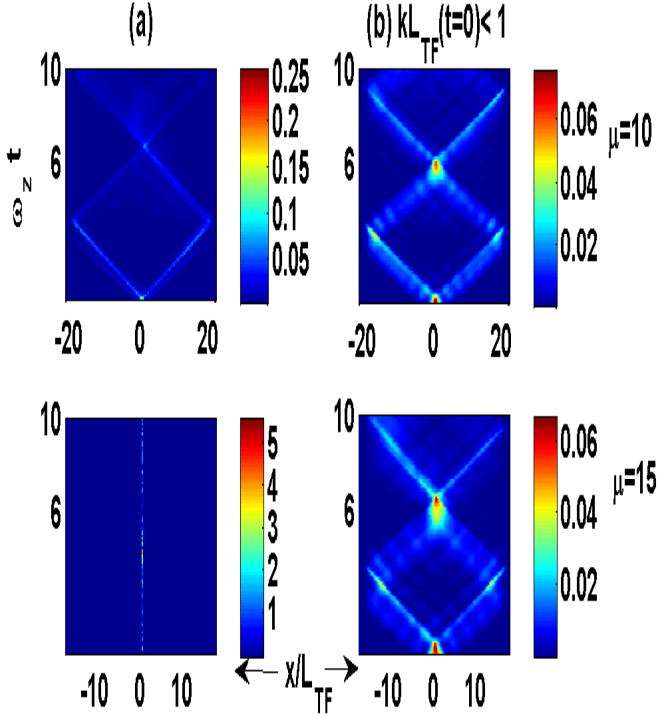}}
\caption{{\it color online}:{The figures in column (a) and (b) gives the full space-time
profile of the density distribution for the same disorder potential used in Fig. \ref{CJSGfig1}. The $x$ and $y$ axes are same for all the plots.
The chemical potentials has been indicated on the right side of each row while the right column again
the initial density profile contains the slow $k$ components satisfying $k L_{TF}(t=0) < 1$.
}}
\label{CJSGfig3}
\end{figure}

In Fig. \ref{CJSGfig3} we have plotted the time evolution of the density profile for the entire time
interval starting from $\omega_z t =0$ to $\omega_z t =10$. From the actual calculation in the spatial grid
we have chosen points at regular interval since  plotting all the points make the size of the file too large to be represented through such color plots. But we have presented the data in a way such that this omission of these intermediate data points does not affect the overall density pattern.  
The upper plot in the left column
for $\mu=10$ ( weaker non-linearity) shows that the cloud
 localizes, expands and then again localizes and does not spread monotonically. 
This is one of the new findings in this paper. 
However we have checked that the edge part of the cloud shows a decaying envelope
 all the time.  At increased non-linearity for $\mu=15$  there is no such oscillation  between spreading and localization and the cloud get
 strongly  localized in the lower plot of the left column. However the corresponding plot in 
 right column where we remove the faster components  from initially such  stronger localization at $\mu=15$ is abscent and we get 
a non-monotonic behavior similar to the one for $\mu=10$. Thus we can argue that the 
lower Fourier components or equivalently the large wavelength
components of the initial density  is primarily responsible for this non-monotonic spreading of the propagating condensate.

To summarize, in the preceding discussion  we have studied the relative impact of non-linearity and disorder on the slow modes
of initial density profile of the  condensate by separating them from the faster ( almost ballistic) moving ones.
We found that non-linearity has much stronger effect on the time evolution as compared to the effect of disorder. The later can 
be enhanced either by lowering $\mu$ or by removing higher $k$ components from the initial condensate density.
Now we extend this analysis for the time evolution of the phase and coherence properties of the condensate. 

\subsection{Phase modulation, uncertainty and coherence properties}
%
\begin{figure}[ht]
\centering
\subfigure[ {\em color online:
Phase (color axis) of the disordered condensate is plotted as a function of space ($\frac{x}{L_{TF}}$,$x$-axis) and time ($\omega_z t$, $y$-axis ) in presence of the disordered potential for $\mu=6$, (left column) and for $\mu=20$ (right column).
$x$ and $y$ axes are same in all the plots. The lower row again gives such phase plot for the corresponding
"slow condensate" for which $kL_{TF}(t=0) <1 $. The disordered potential used is same as one in Fig. \ref{CJSGfig1}}] 
{
    \label{CJSG4a}
    \centerline{{ \epsfxsize 10cm \epsfysize 10cm \epsffile{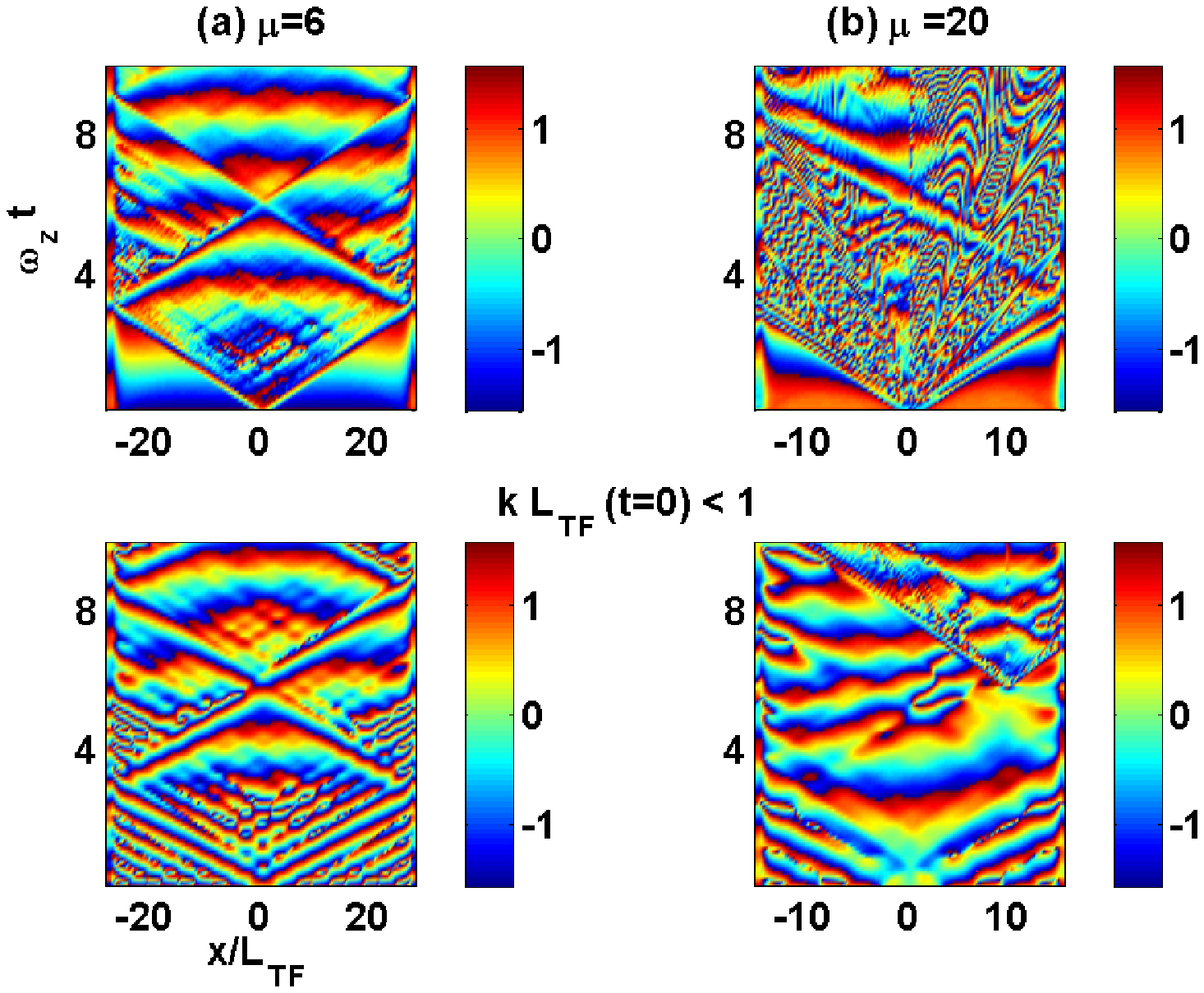}}}
}
\subfigure[{\em color online: Similar space-time plot of the condensate phase for $\mu=6$. Plot (a) is in the absence of  disorder  
and the plot (b) is in the presence of
a stronger disorder than one used in Fig. \ref{CJSG4a} and same as one used in  Fig. \ref{CJSGfig3}.} ] 
{
    \label{CJSG4b}
    \centerline{{\epsfxsize 10cm \epsfysize 5cm \epsffile{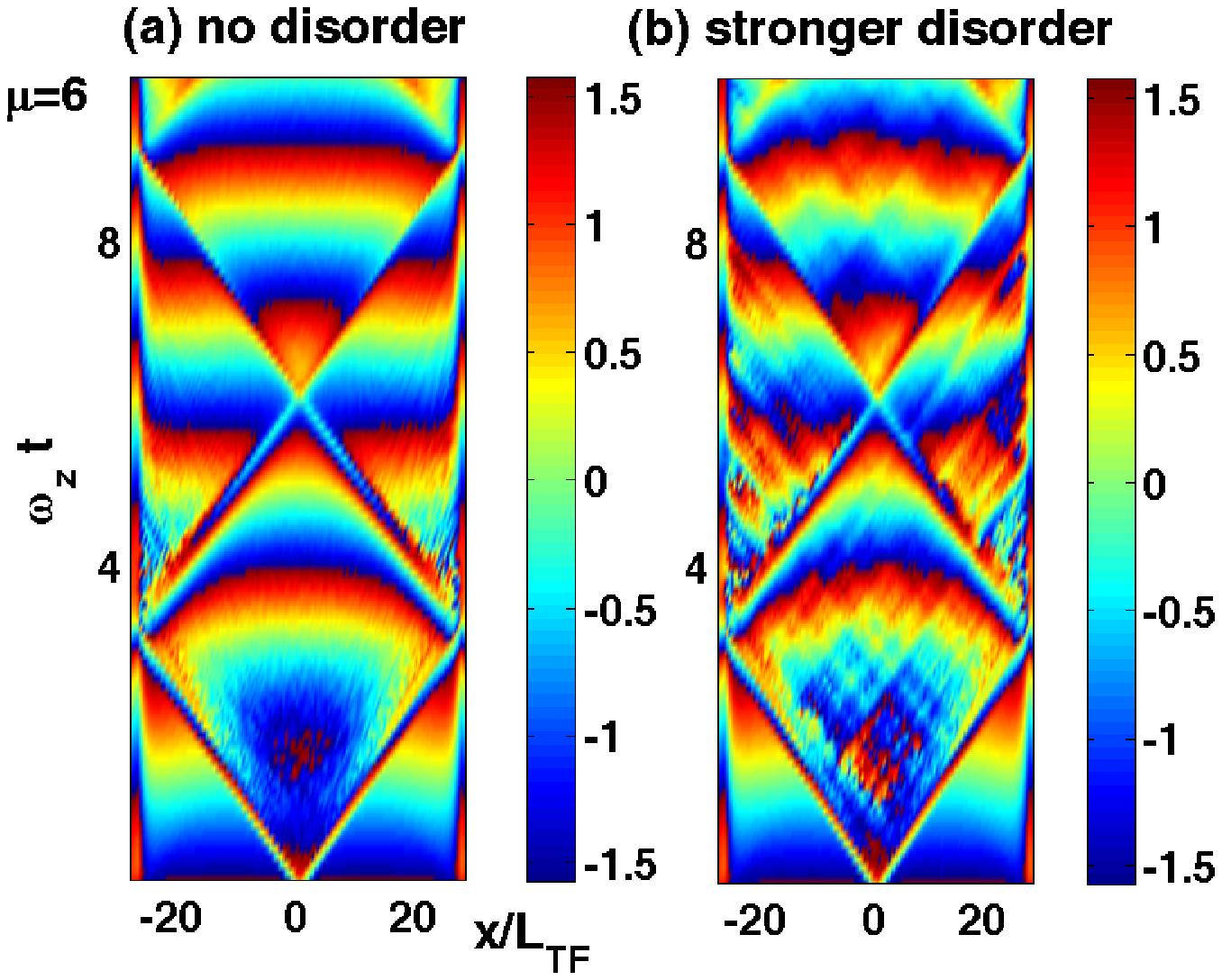}}}
}
\caption{Time evolution of the phase of the propagating condensate with and without disorder}
\label{CJSGfig4} 
\end{figure}
It is known that in addition to the above described time dependent density, during time evolution the condensate attains a dynamical phase \cite{castingora}.
In Fig. (\ref{CJSGfig4}) we plot this  phase of the condensate as a function of position $\frac{x}{L_{TF}}$ and the dimensionless
time $\omega_z t$. The upper row of Fig. \ref{CJSG4a} compares the
space time profile of the condensate phase for a given disorder but varying nonlinearity. As  $\mu$ is increased from $6$ to $20$ to enhance  interaction or non-linearity, the  phase fluctuates on a shorter time and length scale. The details of the phase fluctuation and its degree of non-uniformity in space and time is however dependent on the particular realization of the disorder. For example, in the upper plot of column (b), at later time the phase fluctuation is more on the right side of the origin for the given realization of the
the disorder. It could have happened on the left side as well or at somewhere else for another realization of the disorder potential. The same plot ( first row, second column) also shows that the phase fluctuation is initially concentrated in the center and gradually spreads away. This happens
as the initial profile of the static condensate spread away over a larger region in space with time and happens for any realization of the disorder potential.

The lower row of the Fig. \ref{CJSG4a} gives the corresponding plots for the slow condensate.
The removal of the faster Fourier components
effects the interaction dominated phase fluctuation
( column (b) of the plots) differently as compared to the disorder dominated phase fluctuations (column (a) of the plots). The interaction dominated phase fluctuation gets reduced in this  space-time plot for the slow condensate where as in the left column disorder dominated (lower $\mu$) phase fluctuation gets enhanced for the slow condensate.
However, this change in the fluctuation does not happen uniformly in the space-time. 

In Fig. \ref{CJSG4b} for $\mu=6$, we plot the space time profile of the condensate
phase in the complete absence of the disorder ( left column) and also in the presence of a stronger disorder as compared to the one used 
in Fig. \ref{CJSG4a}. The Fig. \ref{CJSG4b} as well as the upper figure of the left column Fig. \ref{CJSG4a} all of which correspond 
to $\mu=6$, shows that the introduction of the disorder potential and its variation has not any significant effect on the phase fluctuation. 
Thus  increased non linearity has a stronger effect on phase fluctuation
than increased disorder strength. 

Qualitatively, spatial phase fluctuations can be  understood in the following framework. We know after time of order $\omega_z t \sim 1$, most of the interaction energy gets converted into the kinetic energy.
We can write
the condensate wavefunction as $\psi_{w} = \sqrt{\rho_{w}}e^{i \phi_{w}}$, then to a good
approximation ( if the disorder potential is much less than the chemical potential)
\beq i \frac{\partial \psi_{w}}{\partial t} = \mu \psi_{w} ; \mu \approx \frac{1}{4 \rho}(\frac{\partial \rho_{w}}{\partial x})^2 +
 \rho (\frac{\partial \phi_{w}}{\partial x})^2 \nonumber \eeq
Since the typical density is lower than that of the static condensate, 
the most part of this kinetic energy is contributed by the phase fluctuation. Thus we can relate the high phase contrast region with the 
low density regions of the
condensate. We can explain some of the regions with higher or lower phase contrast in this way.  However this simple argument does not explain 
fully all the interesting features of the phase evolution under the combined effect of non-linearity
and disorder.
\begin{figure}[ht]
\centering
\subfigure[ {\em color online:
The uncertainty in position and momentum ($k$) space as a function of time
for two different values of $\mu$  for the disordered condensate. The corresponding $\mu$ has been indicated in each plot. $\Delta x_{SL}$ and $\Delta k_{SL}$ are the corresponding uncertainty in position and momentum space for the slow condensate. The disorder potential is same as one used in 
Fig. \ref{CJSGfig1}.}] 
{
    \label{CJSG5a}
    \centerline{{ \epsfxsize 8cm \epsfysize 7cm \epsffile{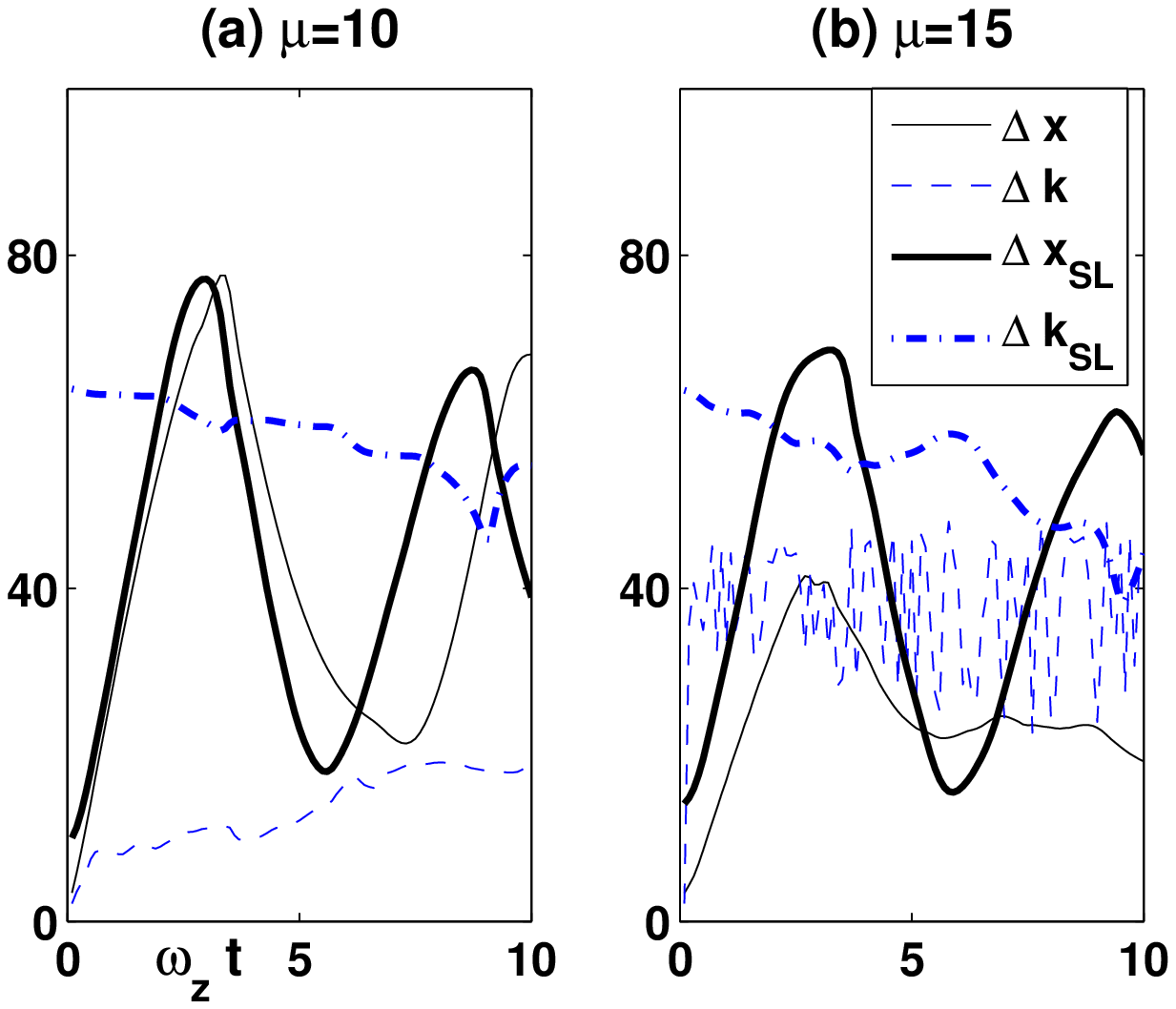}}}
}
\hspace{0.5cm}
\subfigure[{\em  color online: Similar uncertainty plot as in the Fig. \ref{CJSG5a}. $\mu=15$.
Left plot is for the case without any disorder and the right plot is for the case with a stronger disorder.The  stronger
disorder potential  is same as the one used in Fig. \ref{CJSGfig2}. 
The marker used for different plots are same as one used in Fig. \ref{CJSG5a}} ] 
{
    \label{CJSG5b}
    \centerline{{\epsfxsize 8cm \epsfysize 7cm \epsffile{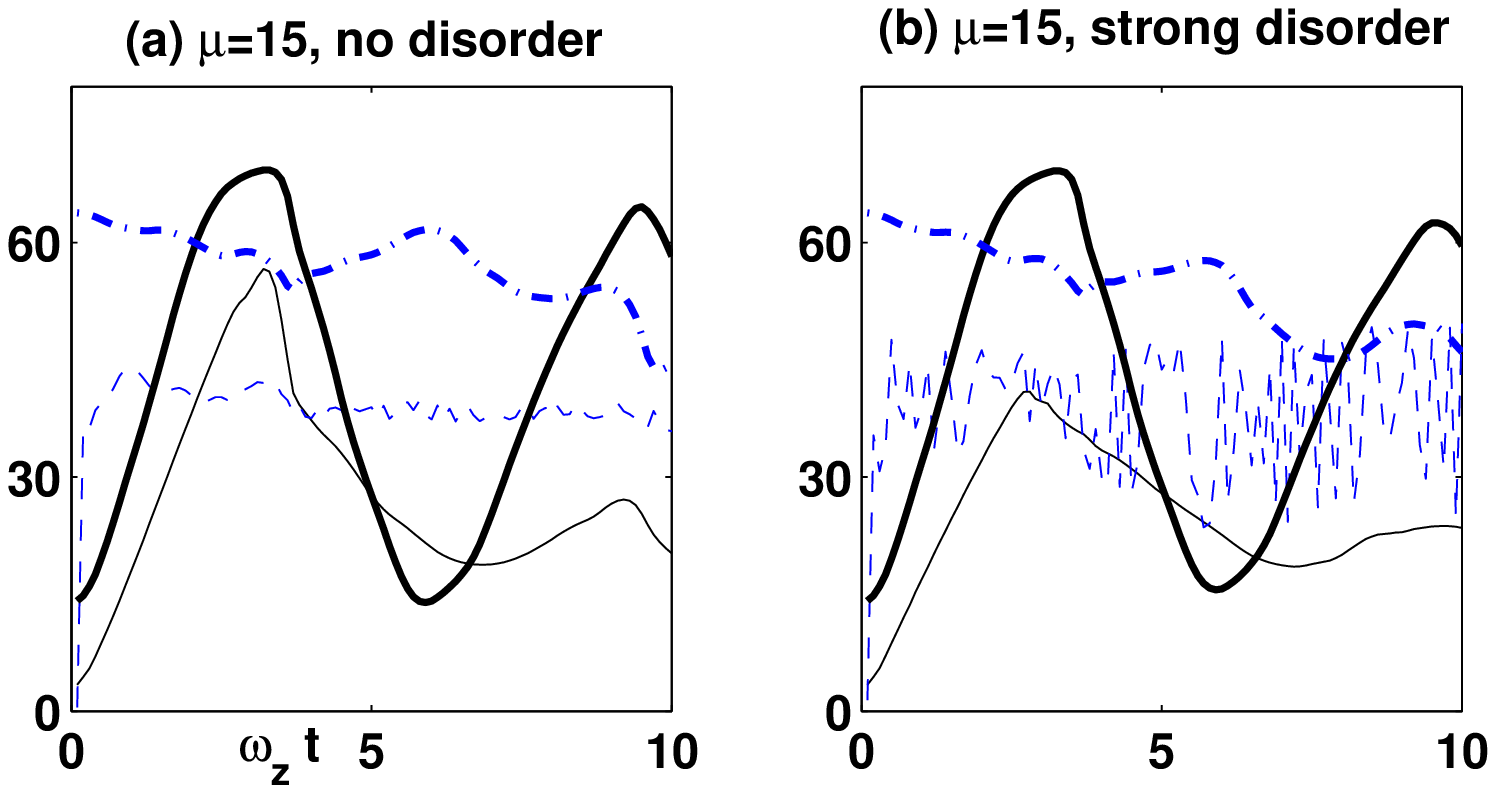}}}
}
\caption{Time evolution of the uncertainty in position and momentum of the condensate}
\label{CJSGfig5} 
\end{figure}
%
\begin{figure}[ht]
\centering
\subfigure[ {\em color online:
Coherence function
$\chi(d,t)$(see Eq. \ref{coherence}) is plotted along the color axis. The $x$ axis is dimensionless time $\omega_z t$ and the $y$ axis is the parameter $d$ in the scale of the $L_{TF}$ of the corresponding condensate.
$\mu$ is  indicated in each upper plot. The disorder potential is same as one used in Fig. \ref{CJSGfig1}. Lower row gives plots for the corresponding slow condensates.}]
{
    \label{CJSG6a}
    \centerline{{ \epsfxsize 9cm \epsfysize 5cm \epsffile{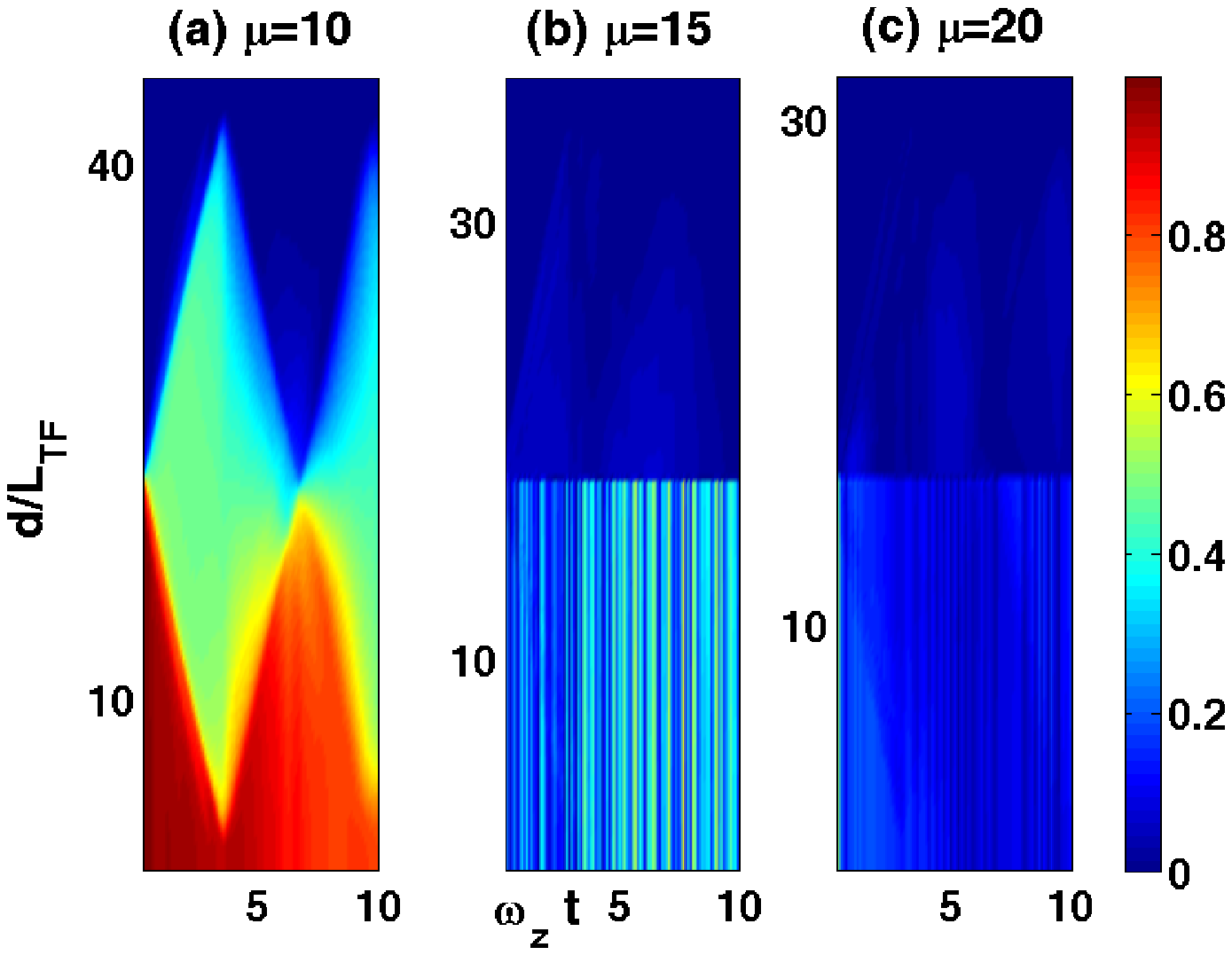}}}
}
\hspace{0.5cm}
\subfigure[{\em  color online: Similar plot of $\chi(d,t)$ as in the Fig. \ref{CJSG6a} for the corresponding slow condensates}]
{
    \label{CJSG6b}
    \centerline{{\epsfxsize 9cm \epsfysize 5cm \epsffile{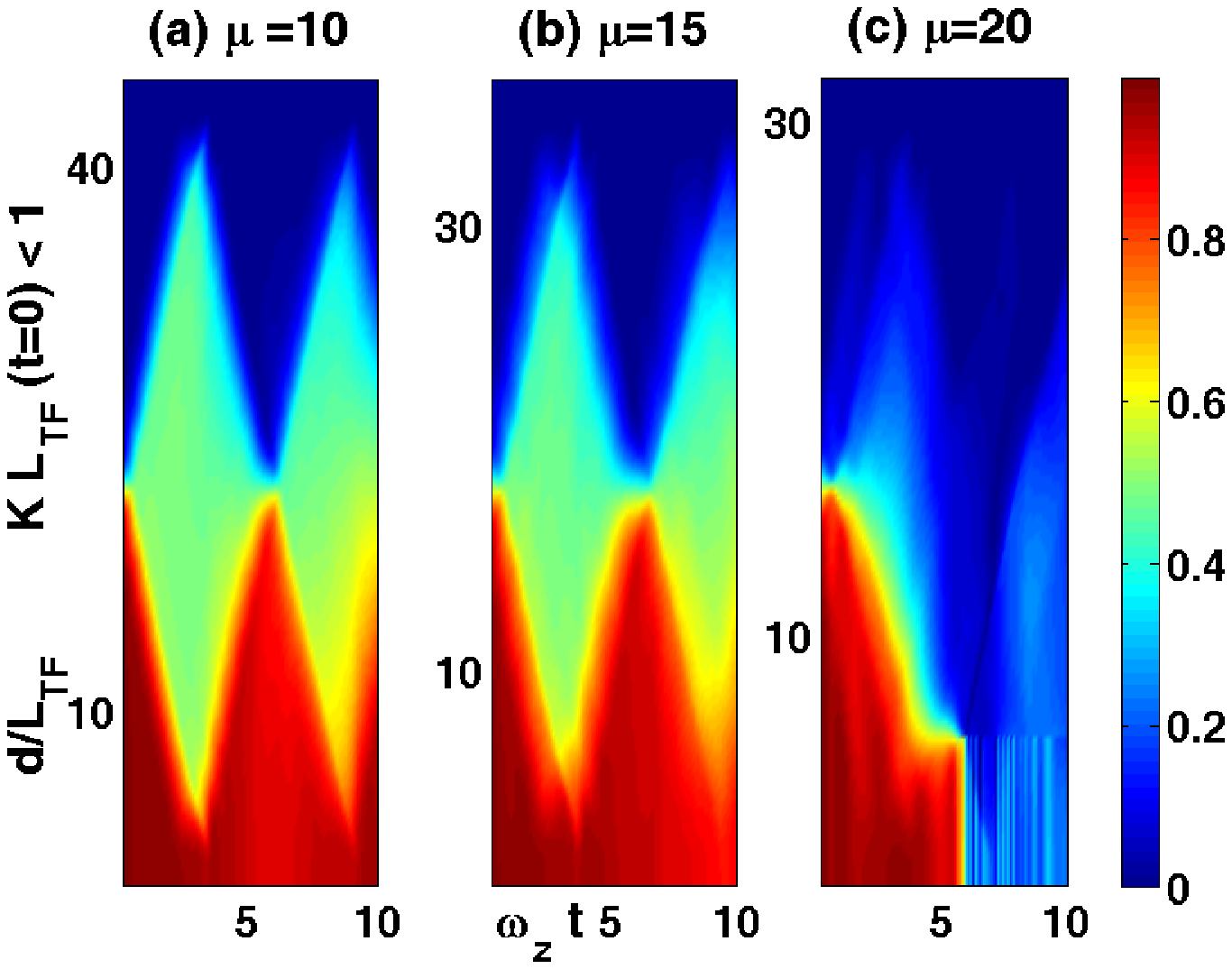}}}
}
\caption{Spatio temporal evolution of the coherence of the condensate for different nonlinearities ($\mu$)}
\label{CJSGfig6} 
\end{figure}

In the next figure  Fig. \ref{CJSGfig5} we plot the uncertainty in the position and the momentum in the cloud. They are defined
in the following way.
\beq \Delta x = \sqrt{\overline{x^2}- \overline{x}^2},~~
\overline{x^n} = \frac{\int dx x^n \rho_\omega (x)}{\int dx \rho_\omega (x)} \, \ \label{spatialextx},
\eeq
where we have characterized the spatial distribution of the cloud by its moments. 
Similarly the  uncertainty in momentum ($k$) can we written as
\beq \Delta k = \sqrt{\overline{k^2}- \overline{k}^2},~~
\overline{k^n} = \frac{\int dk k^n \rho_\omega (k)}{\int dk \rho_\omega (k)} \, \ .\label{spatialextk}
\eeq
where we have characterized the $k$-space distribution of the cloud by its moments,
$\Delta x_{SL}$ and $\Delta k_{SL}$ are the corresponding uncertainty in $x$ and $k$ when the initial condensate is depleted 
of fast Fourier components such that $k L_{TF}(t=0) <1$.
One of the  purposes of these plots is to find out if there is any complementary relation between $\Delta x$ and $\Delta k$ dictated by the uncertainty principle. For example, how much this behavior deviates from a coherent state?
Also the time dependence of  $\Delta x$ or $\Delta x_{SL}$ tells us the nature of the condensate motion. For example, if $\Delta x(t) \sim \sqrt{t}$ the motion is diffusive, if  $\Delta x(t) \sim t$ the motion is ballistic. The saturation of the $\Delta x (t)$ on the other hand
implies inhibition of transport \cite{BECdisorderreview1}.
Thus calculation of $\Delta x(t)$ will provide a direct measure of the impact of non-linearity and disorder on the transport properties of the condensate.

The plots (a) and (b) in Fig. \ref{CJSG5a} are for the same disorder potential that has been used in  Fig. \ref{CJSGfig1}.
Comparing the plots for $\Delta x$ and $\Delta x_{SL}$ in these two figures we find
that the removal of the higher fourier components leads a more rapid temporal change
in uncertainty. This can be clearly seen for a range of  non-linearity ( Fig. \ref{CJSG5a},plots (a) and (b) , $\mu=10,15$). These results agrees with the observation in Fig. \ref{CJSGfig1} where
a strong localization in the special profile in the density for $\mu=15$ is resisted and the condensate is allowed to spread over time in when the higher fourier components are removed from the initial profile of the condensate. Also the oscillatory time
evolutions of  $\Delta x$ in Fig. \ref{CJSG5a} (a) and $\Delta x_{SL}$ in
Fig. \ref{CJSG5a},plots (a) and (b)
are consistent with the time evolution of the density profile given in Fig. \ref{CJSGfig3}. Additionally these figures  show that in 
course of such oscillatory behavior when the condensate spreads, the motion is almost ballistic since it is almost linear in time.


A comparison between the plots Fig. \ref{CJSG5a},plot (b) and Fig. \ref{CJSG5b}, plots (a) and (b)
shows that the increase in the strength of the disorder does not affect the uncertainty
very strongly. Also for  $\mu=15$ (higher non-linearity), $\Delta x$ shows saturation after
some time of flight. We have checked that for even stronger non-linearity (such as $\mu=20$) this saturation takes place in less times as compared to the cases of weaker non-linearity.


Comparing plots of $\Delta k$ and $\Delta k_{SL}$ in
Fig. \ref{CJSG5a}, plot (a) and plot (b) we find that as $\mu$  is being increased, the uncertainty in $k$ shows more fluctuations However we
have checked that at even higher non-linearity ( for example at $\mu=20$) this fluctuation subsides. 
Comparing plots of $\Delta k$ and $\Delta k_{SL}$
in Fig. \ref{CJSG5a},plot (b) and  in Fig. \ref{CJSG5b}, columns  (a) and (b)  we also find that  uncertainty in $k$ fluctuates more with the introduction of the disorder potential.  These fluctuations however subsides for the corresponding slow condensate. Thus we may conclude that the removal of higher fourier components increases the uncertainty in position of the condensate
and reduces the uncertainty in the momentum of the condensate.


\begin{figure}[ht]
\centering
\subfigure[ {\em color online:
Plot of $\chi(d,t)$ (color axis)  as a function of the dimensionless time $\omega_z t$ ($x$-axis)  and  $\frac{d}{L_{TF}}$ ($y$-axis) for $\mu=6$. The upper row is for the condensate. Left plot (a) corresponds
to the case without any disorder  and the right  plot (b) corresponds to the case of stronger disorder potential previously used in
Fig. \ref{CJSG2subb}. }]
{
    \label{CJSG6aa}
    \centerline{{ \epsfxsize 8cm \epsfysize 6cm \epsffile{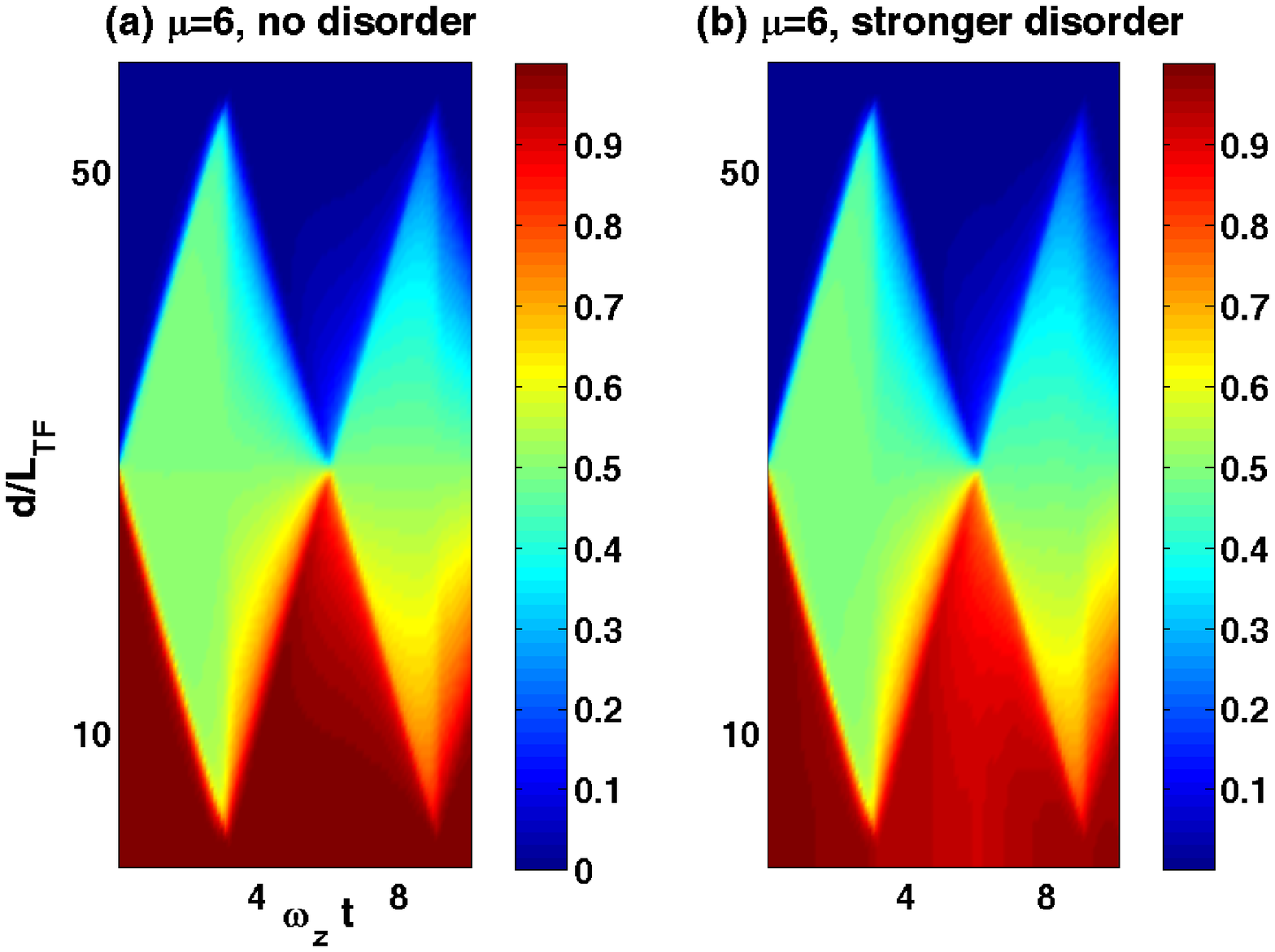}}}
}
\hspace{0.5cm}
\subfigure[{\em  The upper row is a similar plot of $\chi(d,t)$ as in Fig. \ref{CJSG6aa}  but for $\mu=20$. In the lower
we have plotted the corresponding behavior of the slow condensate} ] 
{
    \label{CJSG6ab}
    \centerline{{\epsfxsize 9cm \epsfysize 10cm \epsffile{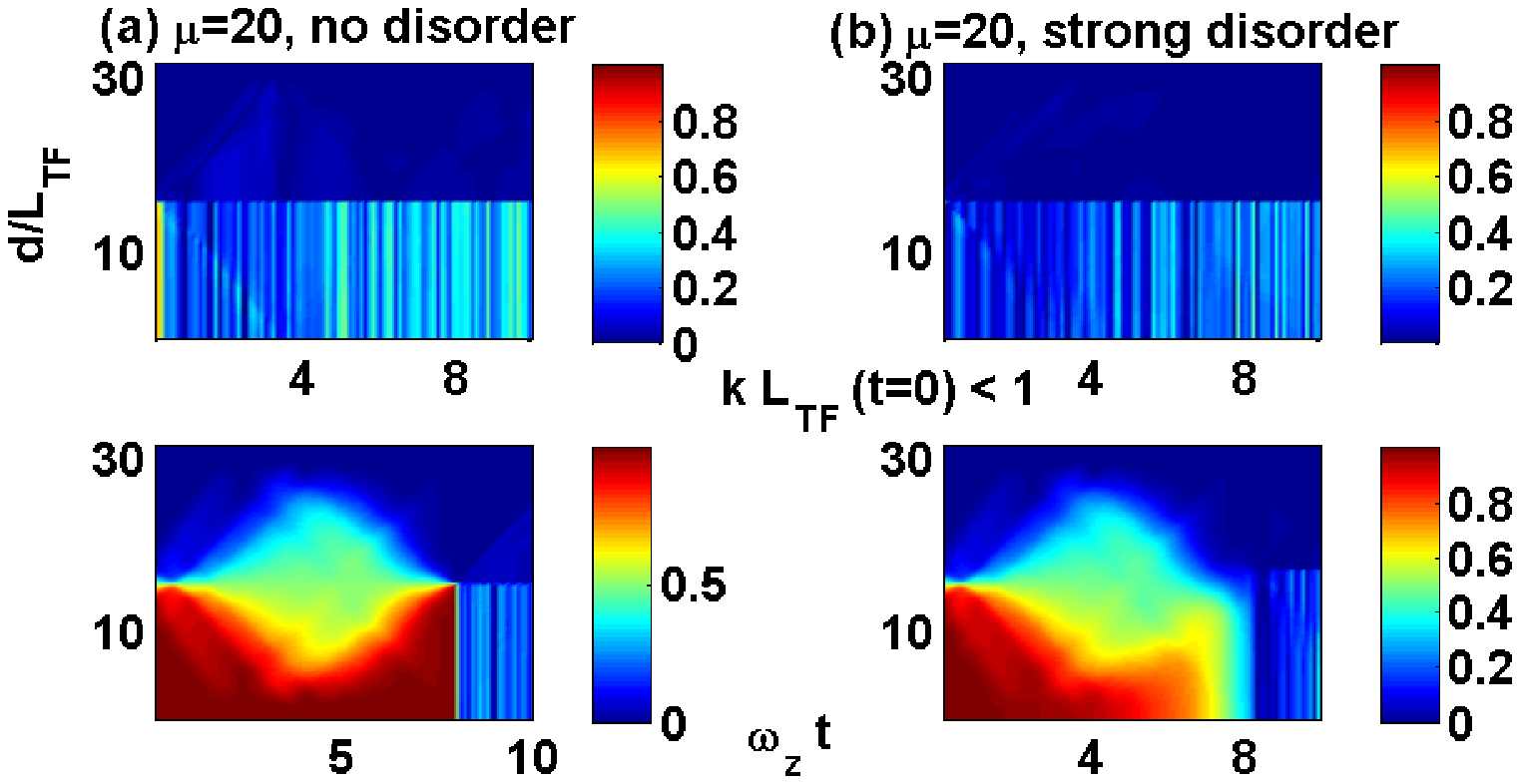}}}
}
\caption{Spatio temporal behavior of coherence in the absence of disorder and in a stronger disorder}
\label{CJSGfig6a} 
\end{figure}
In the next two figures, namely Figs. \ref{CJSGfig6} and \ref{CJSGfig6a}   we plot the coherence in the evolving condensate as a function of space-time. To this purpose following \cite{NM04} we have defined the function 
\beq \chi(d,t)=|\int dx\psi_{\omega}^{*}(x+d,t)\psi_{\omega}(x,t)|^2.  \label{coherence} \eeq
If $\chi(d,t)=1$ there is full phase coherence and if  $\chi(d,t)=0$, the phase coherence is said to be lost. For a given realization of the disorder potential and
 for a given value of $d$, the coherence function is obtained by summing over the product  integrand  in (\ref{coherence}) for all pair of points on the discrete grid which are separated by this  distance $d$ at each time.  If we assume that the ergodic limit is reached this is equivalent to the disorder averaging. We have also numerically verified that features of these plots are common to various realizations of the disorder potential. As the plot shows at a given time coherence changes as a function of the distance between two points. And for a given distance between two points, the coherence changes as a function of time. We shall discuss below the impact of the interaction as well as disorder on this spatio-temporal behavior of this coherence function given in (\ref{coherence})

In Fig. \ref{CJSG6a} we plot such  coherence as a function of space-time  for different strengths of nonlinearity by 
changing $\mu$ but  keeping the disorder potential same. The Fig. \ref{CJSG6b} gives corresponding plots for the slow condensates. 
The Fig. \ref{CJSG6a} shows that for a given separation $d$ between two points on the condensate profile, the coherence decreases with time 
initially and then gets revived again. Such loss and revival of phase coherence has also been reported in numerical simulation of the BEC in a moving optical lattice \cite{NM04}. In Fig. \ref{CJSG6a} as we change from $\mu = 10 \rightarrow 15$, with  
increasing non-linearity, the coherence decreases. Also the oscillation between loss and revival becomes faster ( very thin stripes
of alternating color in the plot (b) of Fig. \ref{CJSG6a}). At a even higher non linearity at  
$\mu=20$, the coherence decays to a very low value at the very initial stage of the time of flight
and the loss and revival behavior also disappears. In Fig. \ref{CJSG6b}, for 
the corresponding slow condensate 
both for $\mu=15$ and $\mu=20$ the coherence is higher as comapred to those in Fig. \ref{CJSG6a}
For $\mu=15$ the loss and revival feature now happens at a slower temporal rate and over a larger region of space. However in Fig. \ref{CJSG6b} for $\mu=20$ 
the coherence decays to a very low value only after $\tau \sim 5$ as compared to Fig. \ref{CJSG6a} and there is no loss of revival 
behavior.These features imply that at a relative high value of chemical potential ($\mu=20$) the phase coherence decays as a function of 
space ($\frac{d}{L_{TF}}$) and time and almost vanishes both in the case  of large separation between two points at all times and also after long time over the entire length of the condensate.

We  contrast this situation with the cases with stronger disorder as well as the case without any disorder
in the next Fig. \ref{CJSGfig6a}. We consider a weakly non-linear case in  Fig. \ref{CJSG6aa} fo $\mu=6$ and strongly non-linear case in Fig. \ref{CJSG6ab} for $\mu=20$.
Comparison between left and right column of both figures shows that the coherence properties are almost  unaffected by disorder irrespective of the strength of the non-linearity. However a comparison between Fig. \ref{CJSG6aa} and Fig. \ref{CJSG6ab} shows that that the coherence function indeed gets affected by the
increase in non-linearity and the time evolved condensate becomes  less coherent
with increasing non-linearity. 


Within our purely numerical framework a complete theoretical understanding of these two intriguing features of coherence function, namely the 
loss and revival of this coherence for moderate non linearity ( such as the case for $\mu=10$ or for slow condensate at $\mu=15$), and its 
independence on the disorder strength, is very difficult.
Thus we confine ourselves in making some relevant comments.
The integrand of the coherence function is dependent on both the density and phase modulation as  $\psi_{\omega} = \sqrt{\rho_{\omega}}e^{i \phi_{\omega}}$. Using this the loss and revival feature can be related to
to the successive spreading and contraction of the condensate density
depicted in Fig. \ref{CJSGfig3}. Generally we find that the coherence
decreases for a spreading condensate  the coherence increases for a contracting condensate. Now let us recall our previous
discussion on phase fluctuation. Since the spreading of the condensate reduces its mean density, the phase fluctuation increases.Increased fluctuation in phase leads to the loss of coherence. Similarly contraction increases the mean density and thus suppresses the phase fluctuations which in turn leads to the gain in coherence. Also from Fig. \ref{CJSG2suba} and Fig. \ref{CJSG2subb} we find that that when the effect of  disorder is enhanced in an otherwise non-linearity dominated bulk region by removing higher Fourier components from the initial profile of the condensate, it just modulates the condensate density according to the modulation of the disorder potential. However the envelope of the density profile does not carry the signature of the disorder except at the edge where the density is very low. Our results suggests that because of this disorder induced modulation the integrand in relation (\ref{coherence}) will be changing very fast, the sum effect of this disorder induced density modulation will be almost zero. However this does not rule out the possibility that the higher order correlation functions involving $\psi_{w}(x,t)$ will also remain insensitive to disorder. 

Thus our plots reveal that the spatio-temporal behavior of the coherence has more to do with non-linearity
rather than disorder. A scattering theory based approach to this problem of loss of superfluidity in presence of disorder has been considered recently in ref. \cite{Paul}.

\subsection{ A brief comparison with a two-dimensional condensate condensate}

\begin{figure}[ht]
\centerline{ \epsfxsize 10cm \epsfysize 8cm
\epsffile{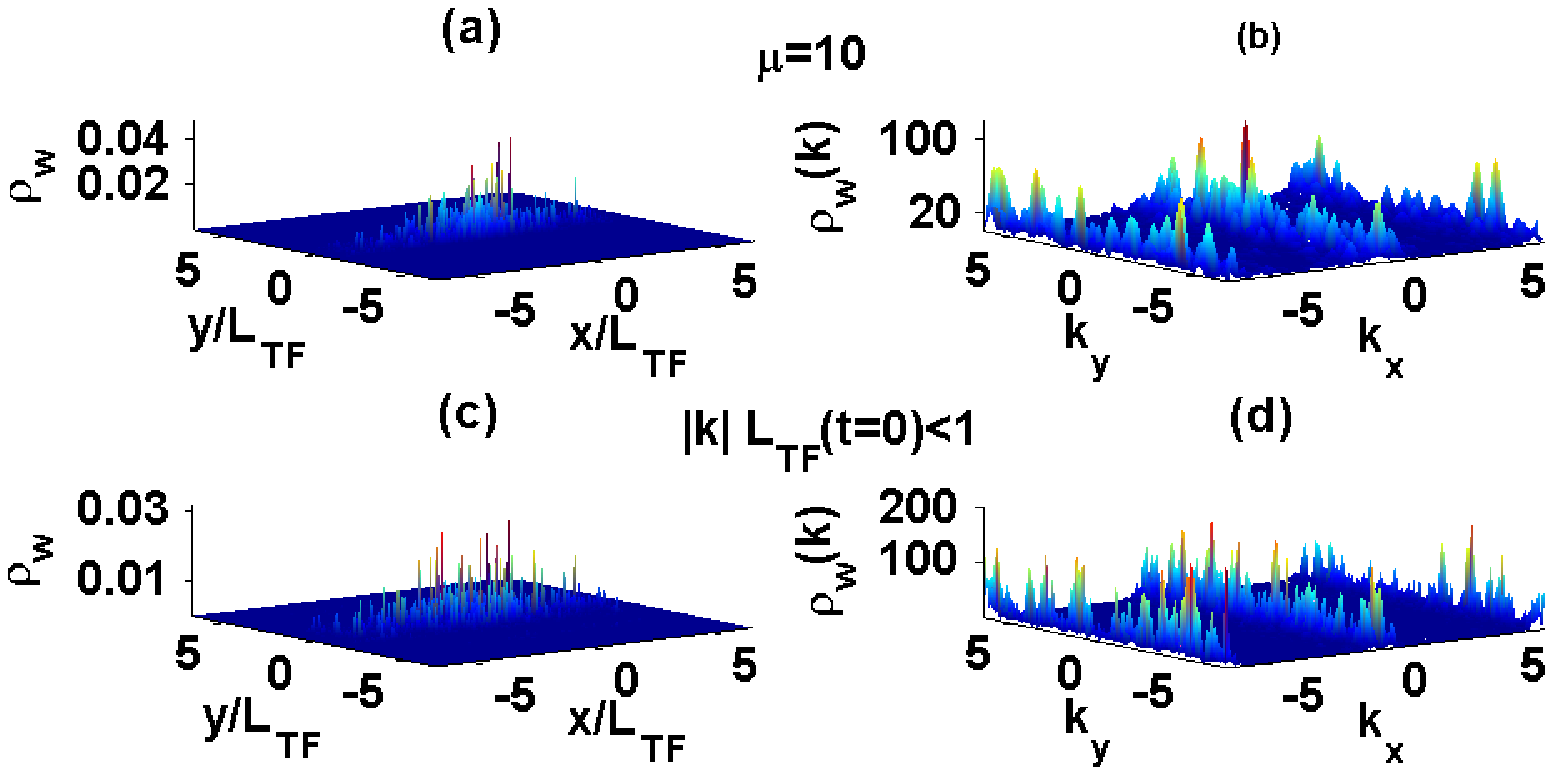}}
\caption{{\it color online}:{ The density plot in co-ordinate and momentum space with $\mu=10$ after dimensionless time $=5$. The disordered potential used here have mean $1.99$ and standard deviation $0.23$.
(a) and (b) gives the density in $x-y$ and $k_x,k_y$ space for the full condensate.
(b) and (d) gives the corresponding density plots after filtering out the higher momentum components
at $t=0$ according to $|k|L_{TF} <1$. Here $L_{TF}$ is actually Thomas Fermi radius in two dimension
and $|k|=\sqrt{k_x^2 + k_y^2}$. The length is measurd in the unit of $L_{TF}$ and $k$ is measured
in the unit of $L_{TF}^{-1}$ }}
\label{CJSGfig7}
\end{figure}

\begin{figure}[ht]
\centerline{ \epsfxsize 10cm \epsfysize 8cm
\epsffile{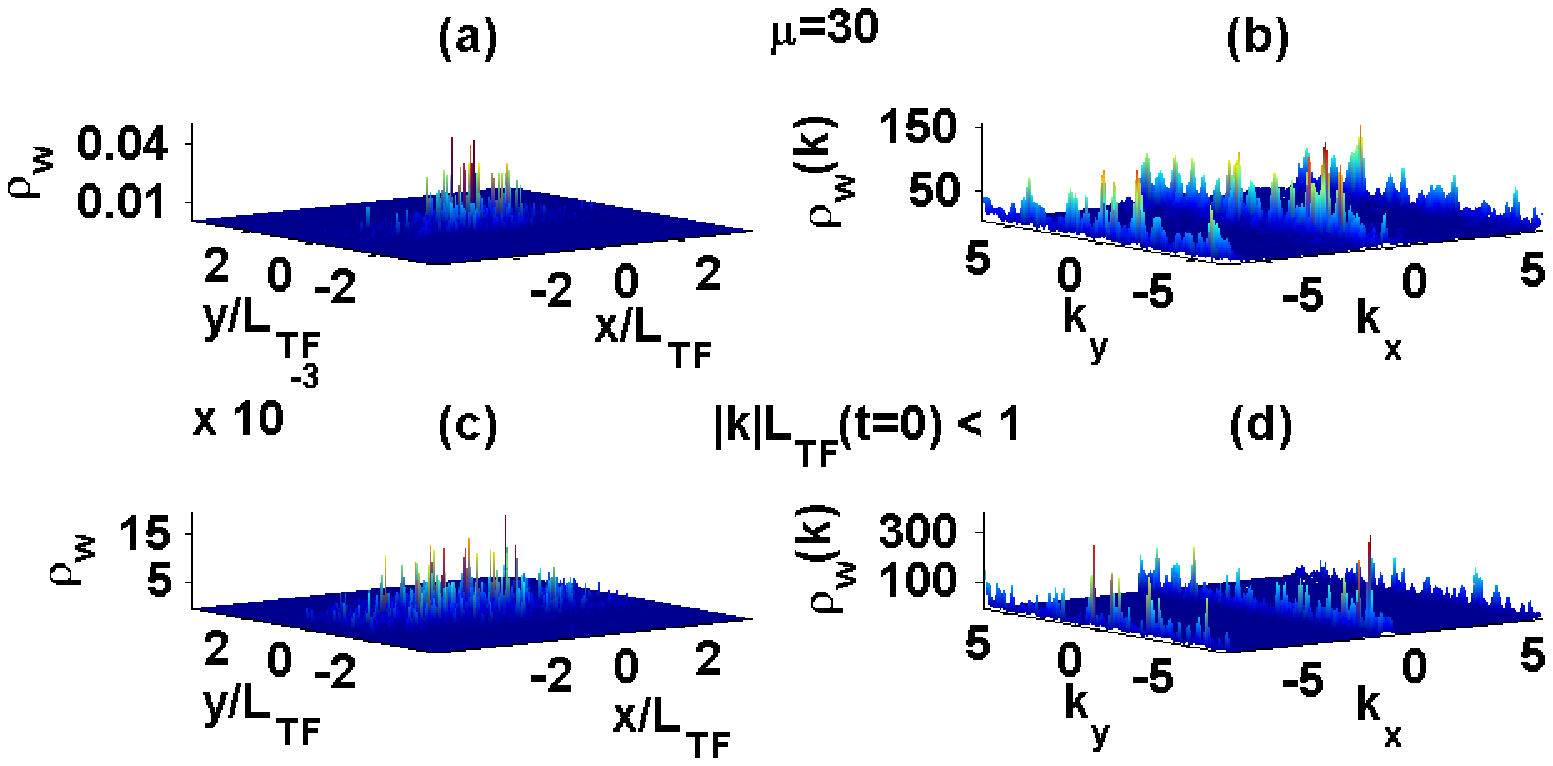}}
\caption{{\it color online}:{ The density plot in co-ordinate and momentum space with $\mu=30$ again after
dimensionless time $=5$. The disorder potential is same as one used in previous figure.
(a) and (b) gives the density in $x-y$ and $k_x,k_y$ space for the full condensate.
(b) and (d) gives the corresponding density plots after filtering out the higher momentum components
at $t=0$ according to $|k|L_{TF} <1$. Here $L_{TF}$ is actually Thomas Fermi radius in two dimension
and $|k|=\sqrt{k_x^2 + k_y^2}$. The length is measured in the unit of $L_{TF}$ and $k$ is measured
in the unit of $L_{TF}^{-1}$ }}
\label{CJSGfig8}
\end{figure}
In this section we compare very briefly the above discussed one dimensional cases with comparable
situations in two dimension. To this purpose, for comparable chemical potentials, we have plotted the density profile of  two-dimensional condensates in a disorder potential after propagating it
for dimensionless time $5$ in position as well as in the momentum space. Also the chemical potential is in the unit of $\hbar \omega_{\perp}$ where $\omega_{\perp}$ is
the confinement frequency in the $x$-$y$ plane.
The results are plotted in Figs. \ref{CJSGfig7} and \ref{CJSGfig8}.
The initial profile
of the condensate is again taken as an Thomas-Fermi condensate with parabolic density profile. Typically the condensate spreads more in two-dimension which agrees with the general assumption that in higher dimension the effect of disorder is less. The another interesting feature is that in the momentum space
density profile we again see a localization of density in the Brillouin zone boundary. Also after the filtering out of the faster components from the initial condensate profile here increases the {\bf spread
of the condensate}, which one can see particularly in the Fig. \ref{CJSGfig8} by comparing the two figures in the first column. A more detailed study of the condensate profile in two dimension in presence of non-linearity and disorder will be presented in a later submission.

\section{Summary of the effect of non-linearity and disorder on the condensate}
The main results of the preceeding discussion can be presented as follows
\begin{itemize}
\item We have tried to differentiate between the effet of disorder and nonlinearity on the time evolution of the condensate 
in a regime where both of them are important. We found that non-linearity always overwhelms the effect of disorder

\item However the effect of non linearity can be reduced either by lowering the chemical potential or by removing faster Fourier components from the initial condensate
    profile

\item If the non linearity is not too strong and the effect of disorder is appreciable 
then the spread of the condensate goes to a non monotonic behavior
by expanding and contracting  during propagation recursively. This behavior is also accompanied by strong 
fluctuation in phase, higher uncertainty in momentum,
and recursive gain and loss in the coherence. With the increase of in the non linearity this behavior is replaced by a strong 
localization of the condensate density in space.
\end{itemize}

%

Thus altogether  our numerical results show a very intriguing non linear dynamics of the density and phase of the condensate when the effect of non-linearity and disorder are present simultaneously .

However all the above results are presented on the basis of  numerical work. The conclusions from
such calculations are always limited due to finite size effect, truncation of the evolution operator, numerical discretization etc.
Nevertheless, they bring forth some very interesting issues and provide some insight for future work.
Particularly the
indication of the non-linear dynamics is intriguing and demands analytical treatment for a complete understanding. As we have pointed out in the beginning that in the successful experimental demonstration
of Anderson localization \cite{ALobservation1} for an interacting BEC, the density profile of the edge
of the condensate is analyzed where effect of the non-linearity is negligible. Here we have mostly focused
on the bulk region where the non-linearity is important to understand its effect on disorder induced quantum localization. Our results will hopefully help to design experiments to probe the effect of non-linearity on Anderson Localization of matter waves.
%
%

\section{Acknowledgement}
SG acknowledges many discussions with Prof. Eric Akkermans  on disordered system in general.
Certain comments and criticism from Prof. Boris Shapiro on the earlier version of the manuscript
are also gratefully acknowledged. SG also acknowledges a very helpful discussion with Prof. T. V. Ramakrishnan on the nature of transport in one dimensional disordered systems.
He also acknowledges financial support from IIT Delhi to attend the Grenoble BEC conference
where a part of this work has been presented.

\end{document}